# Interpreting Pretrained Speech Models for Automatic Speech Assessment of Voice Disorders

Hok Shing Lau[1], Mark Huntly[1], Nathon Morgan[1], Adesua Iyenoma[1], Biao Zeng[2], Tim Bashford[1]

[1]Wales Institute of Digital Information, University of Wales Trinty Saint David, Swansea, UK
[2]Psychology Department, University of South Wales, Pontypridd, UK
`h.lau@uwtsd.ac.uk`

Speech contains information that is clinically relevant to some diseases, which has the potential to be used for health assessment. Recent work shows an interest in applying deep learning algorithms, especially pretrained large speech models to the applications of Automatic Speech Assessment. One question that has not been explored is how these models output the results based on their inputs. In this work, we train and compare two configurations of Audio Spectrogram Transformer [1] in the context of Voice Disorder Detection and apply the attention rollout method [2] to produce model relevance maps, the computed relevance of the spectrogram regions when the model makes predictions. We use these maps to analyse how models make predictions in different conditions and to show that the spread of attention is reduced as a model is finetuned, and the model attention is concentrated on specific phoneme regions.

## 1 Background

### 1.1 Voice for Health

Speech generation involves a complex coordination of organs, as a result, speech contains information about the body from cognitive [3] and mental state [4] to respiration conditions [5] [6]. The decade of research effort into the discovery and utilisation of speech biomarkers, the characteristics of voice in speech associated with certain diseases, for diagnosis, has become more feasible with the advances in Artificial Intelligence (AI) that can learn the association and make clinical predictions. Automatic Speech Assessment, leveraging speech biomarkers, AI and mobile technology to assess patient health remotely, is expected to bring many benefits to early identification and remote monitoring.

A general pipeline of Automatic Speech Assessment involves the following steps: 1). Patient records and uploads speeches in audio waveforms; 2). audio waveforms are pre-processed and Voice parameters, such as fundamental frequency, jitter and shimmer are extracted; 3). a machine learning model such as Support Vector Machines or Decision trees, predicts the clinical outcome. This pipeline can obtain high accuracy [5] in optimal recording conditions but could be less consistent when the underlying voice parameters are impacted by sub-optimal recording conditions, e.g. background noise and device heterogeneity [7] [8].

## 1.2    Deep Learning Speech Model for Automatic Speech Assessment

In recent years, there has been a growing interest in applying deep learning to the problem [9] [10]. This can be done in two approaches, The first approach involves training deep learning models in an end-to-end fashion, such that the model makes clinical predictions directly from the audio, however, proper generalization of the model requires a large amount of manually annotated data, which is time-consuming or not feasible for pathology with smaller sample size.

The second approach involves using a pre-trained deep learning model as a feature extractor and fine-tuning the speech model with lesser annotated data. This type of model is trained on a large corpus of speech data to learn a set of features i.e. representation to capture the attributes of the speech, which then be used in many ASR tasks. It has been shown speech representations capture human perceptual understanding [11] and preserve consistent attributes within the speech such as speaker, language, emotion, and age. As speech contains rich information about the conditions of several important organs, with the rise of these models, there have been several works exploring and evaluating their potential for identifying disease  [12] [13]. However, deep learning models lack interpretability, which hinders their applications in healthcare sectors.

## 1.3    Interpreting speech models

With the rise of concern about AI reliability, tools have been developed to understand models, in general, there are two approaches: white-box and black-box. Black-box approaches systematically probe the model with various tasks and data to estimate its behaviour in a generalised situation, which is known as global explanations, although few approaches such as LIME [14] and SHAP [15] allow local explanations. Black-box approaches are model-agnostic. One example group is perturbation-based methods, as the name suggested, it aims to test model robustness against data perturbation such as adversarial attack or noise, by applying different perturbations.

White-box approaches consist of analysing mathematically the relation between the output and the inputs, therefore, they can provide the local explanation of how output is inferred from the input by the model for a given circumstance. However, they often require specific model architectures and properties such as the existence of activations. In terms of neural networks, there are gradient-based methods that use backpropagation such as Grad-CAM [16] and Integrated Gradient [17], attribute-based such as LRP [18], attention-based such as attention flow and attention rollout [19] [2].

There has been an interest in applying interpretable AI tools to analyse the decision-making of speech models for speech-related applications. Becker et al. present and use AudioMNIST along with LRP to inspect the impact of the input features on the model prediction [20], and have found that the model trained and take spectrogram as inputs focus on lower frequency ranges for sex classification, while models trained on raw waveforms focus on a small fraction of the input. Frommholz et al. use LRP to compare the classification strategies of convolutional neural networks trained on the waveform and spectrogram representations of audio samples for audio event classification [21].

Taking inspiration from interpretable AI in speech applications, the potential benefits of integrating interpretable AI tools into Automatic Speech Assessment are:

1. Understand the underlying decision-making mechanism of the model
2. Guide the design of future models or provide insight into the strategy for improving the model
3. As demonstrated [20], with the proper visualisation, hypotheses about the neural network's feature selection can be made and tested, which can lead to the discovery of clinically relevant voice characteristics.
4. Given any characteristics of the model decisions have been identified, using visualization along with diagnosis will help us determine whether the model decision is reliable

To the best of our knowledge, there has not been work to explore and analyse speech models for Automatic Speech Assessment. In this work, we train and compare two configurations of Audio Spectrogram Transformer [1] in the context of Voice Disorder Detection, and apply the attention rollout method [2] to produce model relevance maps, the computed relevance of the spectrogram regions when the model makes predictions. We use these maps to analyse how models make predictions in different conditions, and how these behaviours of the model change as it is finetuned.

## 2    Methodology

### 2.1    Data selection

The Saarbrücken Voice Database [22] contains recordings from 1002 speakers exhibiting a wide range of voice disorders (454 male and 548 female) and 851 controls (423 male and 428 female). The age of speakers varies from 6-94 years (pathological) and 9-84 years (control). Each recording session contains recordings of /i/, /a/ and /u/ vowels recorded on neutral, higher, lower, rising and falling pitch, and a recording of the short phrase "Guten Morgen, wie geht es Ihnen?". Audio is sampled with professional recording devices at 16-bit 50kHz.

In this work, we group the participants by two categories: gender and pathological status, where pathological status falls into three classes: organic, inorganic and healthy.

The class organic involves physical changes to the tissue, whereas inorganic does not. We follow the approach taken by Huckvale et al. [23] for the selection and classification of pathology, and patients with multiple pathologies that fall into both organic and inorganic will not be selected. We have chosen only the recording of the spoken phrase, all samples were downsampled to 16kHz for the model.

## 2.2 Model Training

In this work, we examine the Audio Spectrogram Transformer AST [1], a Transformer architecture modified from the Vision Transformer [24] and designed for spectrograms. Initially, the input audio waveform of t seconds is padded to the maximum size for the model T seconds and converted into a sequence of 128-dimensional log Mel filterbank (fbank) features computed with a 25ms Hamming window every 10ms, which results in a $128 \times 100T$ spectrogram as input to the AST. The spectrogram is then divided into 16x16 patches and each patch is flattened with a linear projection layer to produce a sequence of embeddings with size 768. Trainable positional embedding (size 768) is added to each of the embeddings to provide the spatial structure of the spectrogram, and class token [CLS] embedding (size 768) is append at the beginning of the sequence, and fed to a transformer encoder. The encoder's output at the class token [CLS] is extracted as the speech representation. The model we use is pretrained on AudioSet [25] and is implemented and available in HuggingFace Transformers [26].

We train the model on the task of binary classification: pathological (organic and inorganic) or healthy subjects. The dataset is stratified and split into train, development and test sets with proportions of 80%, 10% and 10% of the entire dataset. We examine two configurations of the classification models: ast_freeze and ast_finetuned. In ast_freeze, An AST model is set to be non-trainable, and a linear layer is added on top of the model to project the embedding into classification outputs. Construction of ast_finetuned is identical to ast_freeze except the AST model is set to be trainable, and the entire model is fully finetuned. Table *2* shows the training configurations.

## 2.3 Model Decision Interpretation

We implement Chefer et al's attention rollout [2] to visualise model decision-making. The method uses the model's attention layers to produce relevancy maps that visualise the computed relevancy of the spectrogram regions. A relevance map R is initialised with the identity matrix i.e. $R_0 \coloneqq I$ . It uses the attention map $A$ of each attention layer to update the scores on the relevance map. Since each attention map is comprised of multiple heads, the gradients are used to average across heads. The final attention map of a layer $A^-$ is

$$A^- = E_h((\nabla A \odot A)^+)$$

Where $\odot$ is the Hadamard product, $\nabla A \coloneqq \frac{\partial y_t}{\partial A}$, which $y_t$ is the model's output for the class $t$ we visualize, and $E_h$ is the mean across the heads dimension. We follow the original implementation and remove negative contributions before averaging. The final relevance scores on the map $R$ are the relevancy scores aggregated from every layer, the update rule for the relevance scores of a layer $R_l$ is as follows:

$$R_l = R_{l-1} + A^- \cdot R_{l-1}$$

where $R_{l-1}$ is the aggregated relevancy score from the previous layer. To retrieve per-token relevancies for classification tasks, we only take the row corresponding to the class token CLS. The row is then sliced to exclude irrelevant information such as the relevance score for CLS to obtain a sequence with the same dimension as the input for the transformer. To obtain the final relevance map, the sequence is reshaped to the patch's grid size and upsampled back to the size of the original image using bilinear interpolation. Since the model is padded to T second, we truncated the spectrogram to its original time t and normalized it.

As Spectrograms are not natural images, we concatenate the relevance map with the spectrogram into an image, in which the relevance score and spectral power are represented by the hue and brightness of the image respectively. For a better understanding of spectrogram regions, we generated phoneme annotations with the corresponding audio, transcript ("Guten Morgen, wie geht es Ihnen?") and Montreal Force Aligner [27], and added them to the image corresponding to the timestamp.

We manually select samples based on the prediction results of the two models, there are four cases:

- O: where both ast_freeze and ast_finetuned predict correctly,
- X, where both ast_freeze and ast_finetuned predict incorrectly,
- A: where the AST model predicts incorrectly after fine-tuning, i.e. only ast_freeze predicts correctly
- B: where the AST model predicts correctly after fine-tuning, i.e. only ast_finetuned predicts correctly

## 3 Result

### 3.1 Model Performances

**Table 1**: model performance

|  | ast_freeze | ast_finetuned |
|---|---|---|
| Unweighted Average Recall (UAR) | 0.7231 | 0.8199 |
| Area Under the Curve (AUC) | 0.8687 | 0.9106 |

The model performance is shown in Table 1: model performance. We report two measures a). Unweighted Average Recall (UAR), the average recall across all classes without considering the class sample sizes; and b). Area Under ROC Curve (AUC), where the curve measures the model's true-positive rates against the false-positive rates at different classification thresholds. With the based AST model being trainable, the performance of ast_finetuned has noticeably improved.

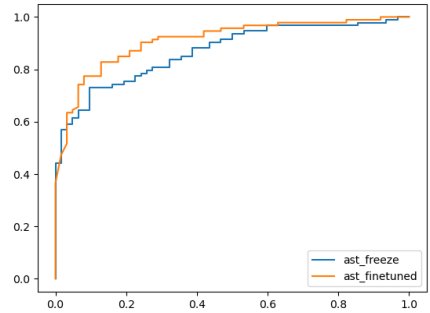

**Figure 1**: receiver operating characteristic curve of two models

### 3.2 Analysis

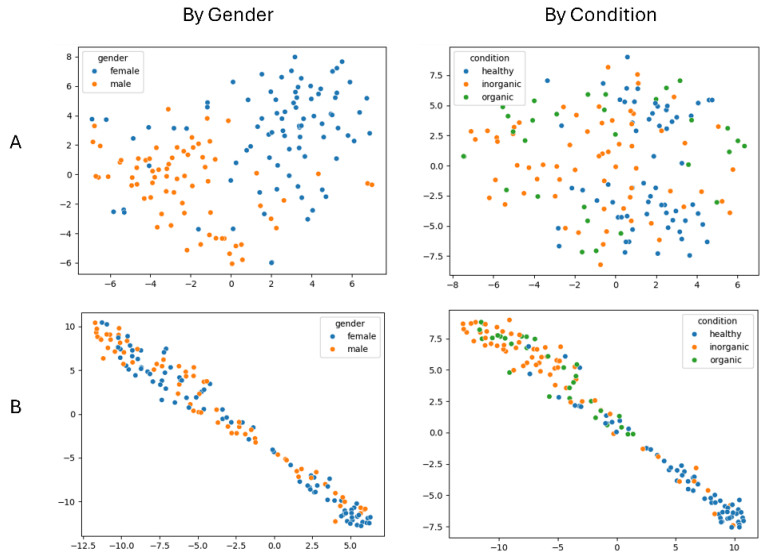

**Figure 2**: t-SNE visualisation of speech representations ast_freeze (top) and ast_finetuned (bottom) by gender (left) and pathological status (right)

We use t-SNE [28] to visualise the speech representations, as demonstrated in Figure 2, when the based AST model is not fully trained (A, ast_freeze), the representations show separations between genders rather than pathological status (pathological vs healthy), in other words, the speech representations contain more information about speakers' gender than the status regarding the potential voice pathologies. On the other hand, it shows the opposite trend when based AST model is fully trained (B, ast_fine-tuned). None of the models shows a clear separation between organic and inorganic pathologies.

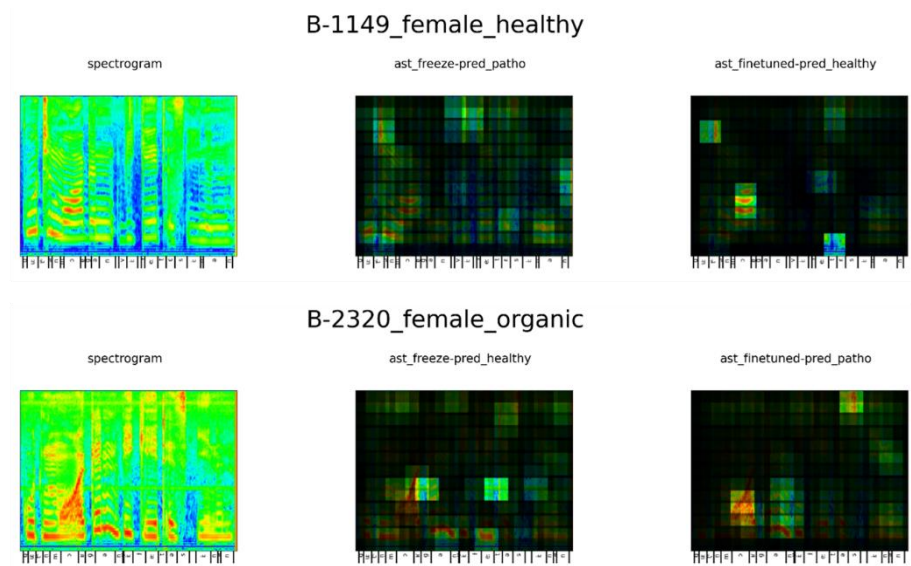

**Figure 3**: Spectrogram (left) and Relevance Map of ast_freeze (middle) and ast_finetuned (right) for two female speech samples (top: healthy, bottom: pathological): The prediction results of these two samples are labelled as B i.e. ast_finetuned predicted correctly whereas ast_freeze incorrectly.

Figure 3 shows two of the example visualizations, more are available in the appendix. From the available visualizations, we can see the highest relevance scores are not necessarily assigned to the highest-intensity region such as fundamental frequency and the formats. A more common pattern that appeared on both of the models is that they give higher scores to phoneme "/ɔ/" and segment "/e/ /s/ /i/ /n/". When the model was fine-tuned, we found more concentration, and the location often changed/shifted, however, no noticeably consistent patterns were concluded.

Traditional voice parameters offer more explainability than the deep learning model, but iterative experiments are required to find the optimal feature selections. Since a lot of voice parameters are derived from spectrogram representations, the relevance map of the spectrogram allows us to prioritise the selection of features.

There are models pre-trained using the phonetic aspects of speech, such as [29], but the AST models we examine are not trained with this in mind. However, the visualizations demonstrate that both models tend to focus on certain phonemes when performing Automatic Speech Assessment. This result matches and seems to support the idea that Automatic Speech Assessment benefits from specific phonetic content [30] [31], which requires consideration and design of speech tasks.

## 4    Conclusion

In this work, we train and compare two configurations of Audio Spectrogram Transformer [1] in the context of Voice Disorder Detection and apply the attention rollout method [2] to produce the relevance map of the models. We use these maps to analyse the computed relevancy of the spectrogram regions. Through the analysis, we found the models cannot fully identify the difference between organic and inorganic voice disorders, models give higher scores to the phoneme "/ɔ/" and segment "/e/ /s/ /i/ /n/". While the model is fine-tuned, we found that the spread of attention has often been reduced. These findings demonstrate the potential importance of phoneme features for Automatic Speech Assessment.

As a follow-up work, we will explore and compare how the model decision-making changes when performing organic vs non-organic using the same data selection and models, and we will compare the results with [30]. One of the issues we did not address in this work is the impact of recording environments on model decision-making, which is important to model robustness under suboptimal environments.

In the future, we will explore voice parameters and develop speech tasks that match the pattern on the relevance map of the spectrogram, we plan to conduct experiments in other pathology speech databases to test whether these features and speech tasks allow better performance. We will also conduct a similar analysis on other models that are pre-trained using the phonetic aspects of speech, which would likely influence the selection of the interpretation method.

**Acknowledgments.** The authors would like to acknowledge the support of the Wales Institute for Digital Information and University of Wales Trinity Saint David in completing this work.

**Disclosure of Interests.** The authors declare that there are no competing interests relevant to this work.

# 5 Appendix

## 5.1 Model configuration

**Table 2**: Model configurations of two examined models ast_freeze and ast_finetuned in Huggingface transformers implementation

|  | ast_freeze | ast_finetuned |
|---|---|---|
| model_class | ASTForAudioClassification | ASTForAudioClassification |
| model_path | MIT/ast-finetuned-audioset-10-10-0.4593 | MIT/ast-finetuned-audioset-10-10-0.4593 |
| num_labels | 2 | 2 |
| freeze | TRUE | FALSE |
| evaluation_strategy | epoch | epoch |
| save_strategy | epoch | epoch |
| learning_rate | 0.001 | 0.00025 |
| per_device_train_batch_size | 8 | 8 |
| gradient_accumulation_steps | 4 | 4 |
| per_device_eval_batch_size | 8 | 8 |
| num_train_epochs | 10 | 40 |
| warmup_ratio | 0.1 | 0.1 |
| logging_steps | 50 | 50 |
| eval_steps | 50 | 50 |
| push_to_hub | FALSE | FALSE |
| remove_unused_columns | FALSE | FALSE |
| early_stopping_patience | 5 | 8 |
| early_stopping_threshold | 0 | 0 |

## 5.2 Visualisations

Figures in this section are Spectrogram (left) and the Relevance Map of ast_freeze (middle) and ast_finetuned (right) for speech samples, title follows the naming convention: "**PredictionResults-Speaker ID_Gender_PathologicalStatus**", there are four cases for **PredictionResults**:

- O: where both ast_freeze and ast_finetuned predict correctly,
- X, where both ast_freeze and ast_finetuned predict incorrectly,
- A: where the AST model predicts incorrectly after fine-tuning, i.e. only ast_freeze predicts correctly
- B: where the AST model predicts correctly after fine-tuning, i.e. only ast_finetuned predicts correctly

The small title follows the naming convention: "**Model**-pred_**Prediction**", there are two classes for **Prediction**: healthy and patho(logical).

### A-87_male_healthy

spectrogram     ast_freeze-pred_healthy     ast_finetuned-pred_patho

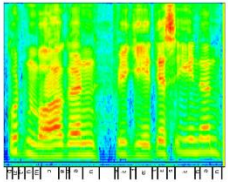 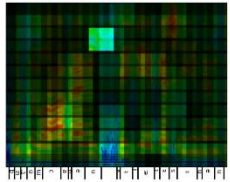 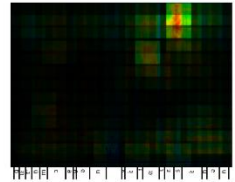

### A-1289_male_inorganic

spectrogram     ast_freeze-pred_patho     ast_finetuned-pred_healthy

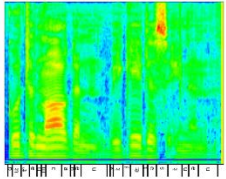 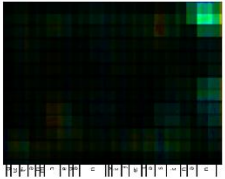 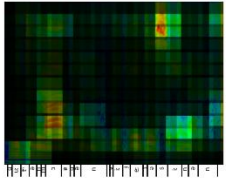

### A-1750_female_inorganic

spectrogram     ast_freeze-pred_patho     ast_finetuned-pred_healthy

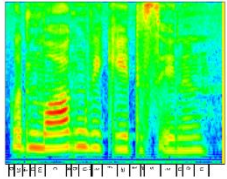 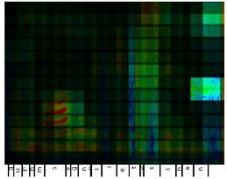 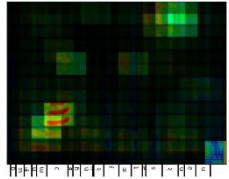

### B-681_male_healthy

spectrogram     ast_freeze-pred_patho     ast_finetuned-pred_healthy

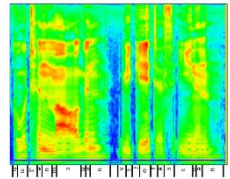 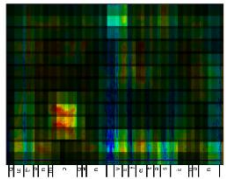 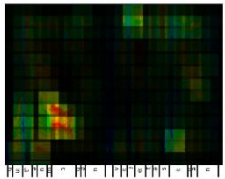

### B-1149_female_healthy

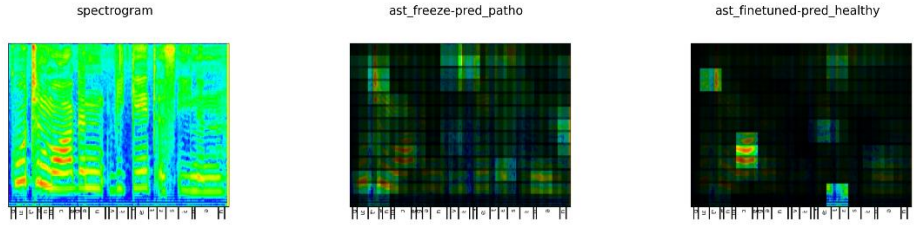

### B-2320_female_organic

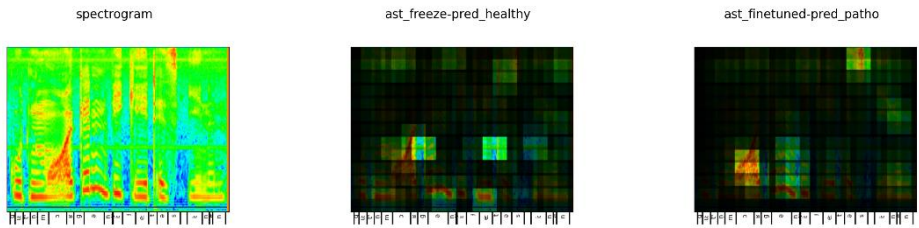

### O-1506_female_healthy

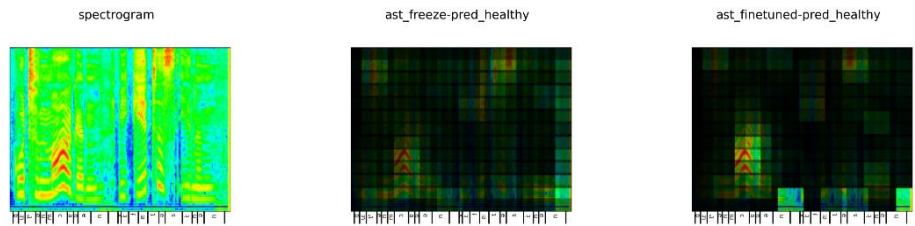

### O-1756_male_inorganic

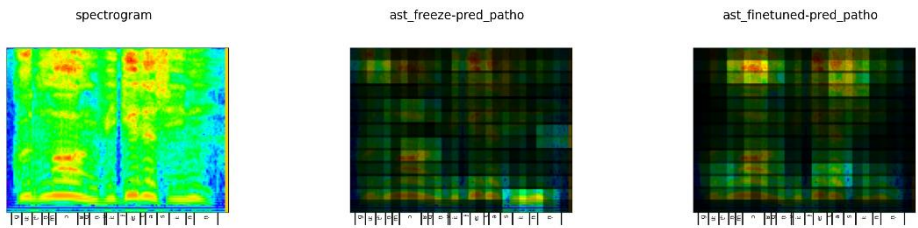

### O-2553_female_inorganic

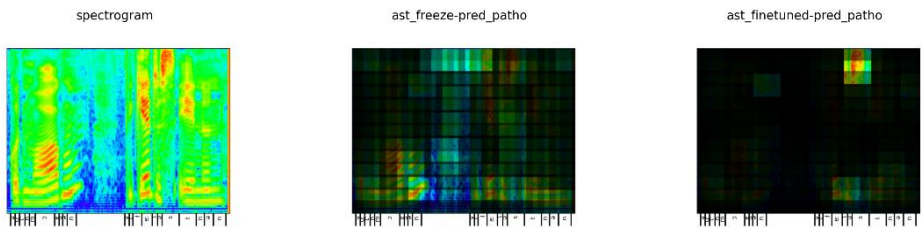

### O-2580_male_organic

spectrogram       ast_freeze-pred_patho       ast_finetuned-pred_patho

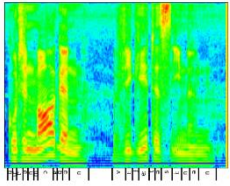 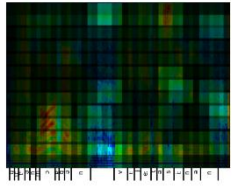 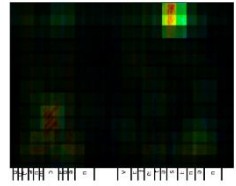

### X-983_male_healthy

spectrogram       ast_freeze-pred_patho       ast_finetuned-pred_patho

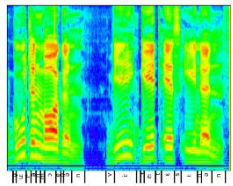 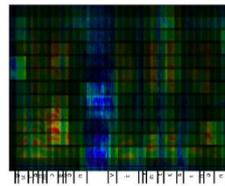 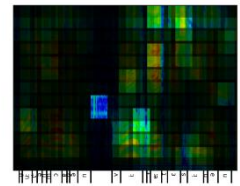

### X-1178_female_healthy

spectrogram       ast_freeze-pred_patho       ast_finetuned-pred_patho

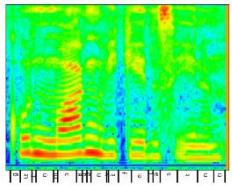 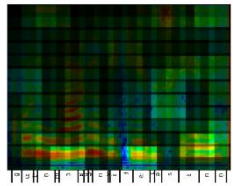 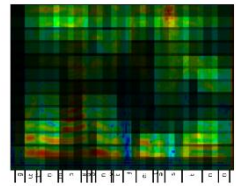

### X-1766_female_inorganic

spectrogram       ast_freeze-pred_healthy       ast_finetuned-pred_healthy

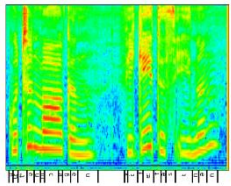 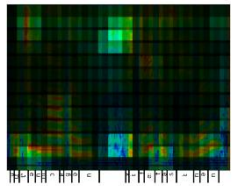 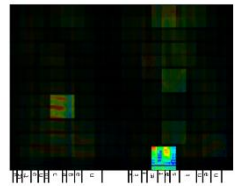

### X-2289_male_inorganic

spectrogram       ast_freeze-pred_healthy       ast_finetuned-pred_healthy

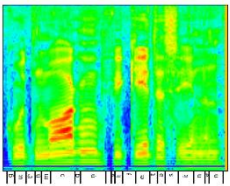 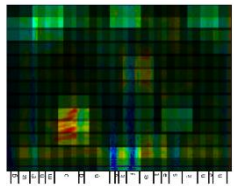 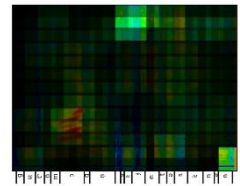